\documentclass[conference]{IEEEtran}
\IEEEoverridecommandlockouts
\usepackage{cite}
\usepackage{amsmath,amssymb,amsfonts}
\usepackage{algorithmic}
\usepackage{graphicx}
\usepackage{textcomp}
\usepackage{xcolor}
\usepackage{lipsum} 

\usepackage{filecontents}

\def\BibTeX{{\rm B\kern-.05em{\sc i\kern-.025em b}\kern-.08em
    T\kern-.1667em\lower.7ex\hbox{E}\kern-.125emX}}
\begin{document}

\title{Uncertainty-Aware, Risk-Adaptive Access Control for Agentic Systems using an LLM-Judged TBAC Model}

\author{\IEEEauthorblockN{ Charles Fleming}
\IEEEauthorblockA{ 
\textit{Cisco Systems}\\
San Jose, USA \\
chflemin@cisco.com}
\and
\IEEEauthorblockN{ Ashish Kundu}
\IEEEauthorblockA{ 
\textit{Cisco Systems}\\
San Jose, USA \\
ashkundu@cisco.com}
\and
\IEEEauthorblockN{Ramana Kompella}
\IEEEauthorblockA{ 
\textit{Cisco Systems}\\
San Jose, USA \\
rkompell@cisco.com}
}

\maketitle

\begin{abstract}
The proliferation of autonomous AI agents within enterprise environments introduces a critical security challenge: managing access control for emergent, novel tasks for which no predefined policies exist. This paper introduces an advanced security framework that extends the Task-Based Access Control (TBAC) model by using a Large Language Model (LLM) as an autonomous, risk-aware judge. This model makes access control decisions not only based on an agent's intent but also by explicitly considering the inherent \textbf{risk associated with target resources} and the LLM's own \textbf{model uncertainty} in its decision-making process. When an agent proposes a novel task, the LLM judge synthesizes a just-in-time policy while also computing a composite risk score for the task and an uncertainty estimate for its own reasoning. High-risk or high-uncertainty requests trigger more stringent controls, such as requiring human approval. This dual consideration of external risk and internal confidence allows the model to enforce a more robust and adaptive version of the principle of least privilege, paving the way for safer and more trustworthy autonomous systems.
\end{abstract}


\section{Introduction}

The advent of sophisticated AI agents marks a pivotal moment in digital transformation, but their capacity for \textbf{emergent behavior} poses a fundamental challenge to traditional security models. Existing frameworks like Role-Based Access Control (RBAC) and Attribute-Based Access Control (ABAC) are ill-equipped to handle novel tasks for which no policies exist. The consequence of a flawed access decision is magnified when the actor is an autonomous agent capable of executing complex actions at machine speed. While using a Large Language Model (LLM) to interpret agent intent and generate policies is a promising direction, it introduces two new critical questions. First, how does the system account for the inherent \textbf{risk} of the resources an agent wishes to access? A request to read a test database is fundamentally different from a request to modify a production firewall. Second, how can we trust the LLM's judgment? An LLM may generate a syntactically correct but logically flawed plan with a high degree of confidence, failing to recognize its own knowledge gaps.

This paper argues that for an autonomous authorization system to be trustworthy, it must be both risk-aware and self-aware. To this end, we propose an \textbf{Uncertainty-Aware, Risk-Adaptive TBAC} model. This framework enhances the LLM Judge by requiring it to explicitly reason about two additional dimensions:
\begin{enumerate}
    \item \textbf{Resource Risk:} Each tool, API, or data source is assigned a risk score based on its criticality.
    \item \textbf{Model Uncertainty:} The LLM provides a confidence estimate for its generated plan, quantifying its certainty in the proposed course of action.
\end{enumerate}
The final authorization decision is a function of the task's intent, its composite risk score, and the LLM's uncertainty. This creates a dynamic, multi-faceted control plane that can distinguish between a low-risk, certain request and a high-risk, uncertain one, enabling a truly adaptive and robust implementation of the principle of least privilege. This paper proceeds as follows: Section II reviews related work. Section III defines the foundational TBAC model. Section IV introduces the extension of TBAC using an LLM Judge. Section V details our proposed uncertainty-aware, risk-adaptive framework. Section VI provides use-case analyses. Section VII discusses practical implementation challenges, and Section VIII concludes the paper.

\begin{figure*}[h]
    \centering
    \includegraphics[width=0.75\linewidth]{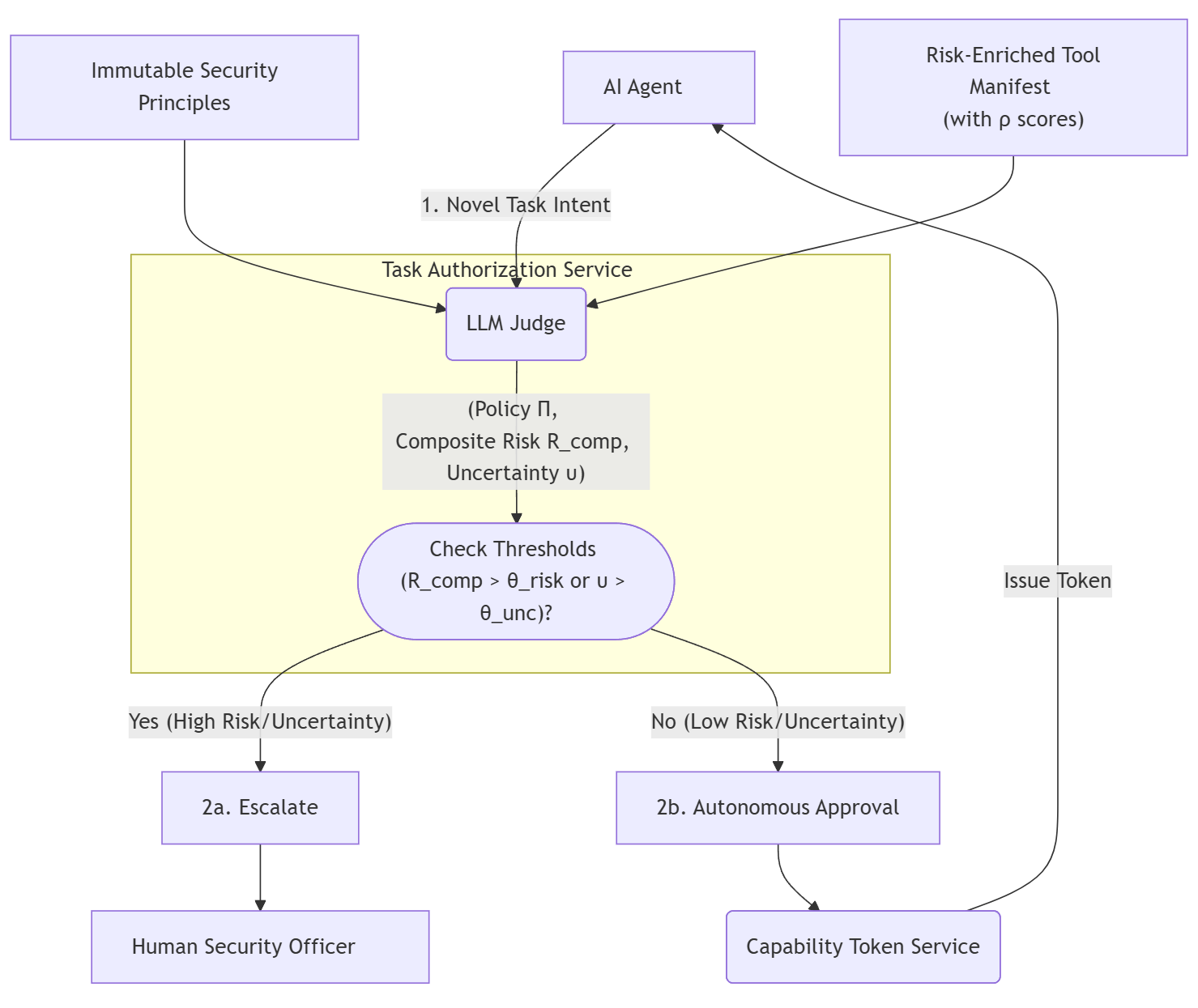}
    \caption{High-level architecture of the proposed TBAC model. The LLM Judge within the Task Authorization Service (TAS) synthesizes a policy while also assessing task risk and its own uncertainty, feeding a dynamic decision engine.}
    \label{fig:main_arch}
\end{figure*}

\section{Background and Related Work}

Our work is situated at the intersection of access control theory, risk management, and uncertainty quantification in machine learning.

\subsection{Evolution of Access Control Models}
Access control has evolved from static, identity-centric models to more dynamic, context-aware paradigms. \textbf{Role-Based Access Control (RBAC)} \cite{sandhu1996role} simplified administration by assigning permissions to roles rather than individual users. However, RBAC struggles with dynamic environments where access needs cannot be neatly categorized by job function. \textbf{Attribute-Based Access Control (ABAC)} \cite{hu2014guide} offered more granularity by evaluating policies based on attributes of the user, resource, and environment. While flexible, ABAC requires authoring complex policies and does not inherently understand the "intent" behind an access request.

\textbf{Task-Based Access Control (TBAC)} \cite{thomas1997task} and related workflow-based models \cite{atluri1996authorization} represented a shift toward intent-driven authorization. In TBAC, permissions are bundled into tasks, ensuring users receive the minimal set of privileges needed for a specific activity, for the duration of that activity. Our work uses TBAC as its foundation due to this natural alignment with the goal-oriented nature of AI agents.

\subsection{Risk and Uncertainty in Access Control}
The concept of incorporating risk into authorization is not new. \textbf{Risk-Adaptive Access Control (RAAC)} models \cite{das2017adaptive} dynamically adjust permissions based on real-time risk assessments. These models often calculate risk based on contextual factors like user location, device posture, time of day, or behavioral anomalies. However, they typically rely on predefined rules and do not address the challenge of assessing the risk of novel, agent-generated tasks based on the composition of resources they intend to use.

Simultaneously, the field of machine learning has developed robust techniques for \textbf{quantifying model uncertainty}. It is well understood that deep learning models can be unreliable when faced with out-of-distribution data. This uncertainty is often categorized as either \textit{aleatoric} (inherent randomness in the data) or \textit{epistemic} (due to model limitations). Our work focuses on epistemic uncertainty. Techniques like Bayesian approximation using Monte Carlo dropout \cite{gal2016dropout} or conformal prediction \cite{angelopoulos2021gentle} provide statistically grounded methods for a model to express "I don't know." Applying this concept to an LLM Judge is critical; an LLM that can communicate its own uncertainty is inherently safer than one that is always confidently wrong. Our model is the first, to our knowledge, to synthesize these two lines of research—task-based risk assessment and model uncertainty—into a cohesive framework for securing autonomous agents.

\section{Task-Based Access Control: The Foundational Model}

The core of our system is rooted in the Task-Based Access Control (TBAC) model. Unlike models centered on user roles, TBAC grants permissions based on the specific tasks an agent needs to perform. This intent-driven approach is a natural fit for agentic systems.

\subsection{Formal Model Definition}
In a standard TBAC framework, we define the core entities:
\begin{itemize}
    \item A set of agents, $A = \{a_1, a_2, \dots\}$.
    \item A set of tasks, $T = \{t_1, t_2, \dots\}$.
    \item A set of available tools (or resources), $S_{tool} = \{s_1, s_2, \dots\}$.
    \item A set of possible transactions on tools, $S_{trans} = \{\text{read}, \text{write}, \text{execute}, \dots\}$.
\end{itemize}
An authorization for an agent $a \in A$ to perform a task $t \in T$ is a policy, $\Pi$. This policy is the set of permissions required to complete the task:
$$ \Pi = \{ (s, tx) | s \in \tau_{tools}(t), tx \in \tau_{trans}(t, s) \} $$
where $\tau_{tools}: T \rightarrow 2^{S_{tool}}$ is a function mapping a task to the set of tools it needs, and $\tau_{trans}: T \times S_{tool} \rightarrow 2^{S_{trans}}$ is a function mapping a task and a tool to the set of required transactions. This authorization is granted at the start of task $t$ and is valid only for its duration.

\subsection{Reference Architecture and Limitations}
The traditional TBAC architecture consists of three main components:
\begin{itemize}
    \item \textbf{Task Authorization Service (TAS):} A central entity that evaluates a task request and grants a policy.
    \item \textbf{Capability Token Service (CTS):} Issues a short-lived, cryptographically signed token (e.g., a JWT) that encodes the policy $\Pi$.
    \item \textbf{Policy Enforcement Points (PEPs):} Services or gateways that protect resources. They intercept requests, validate the token from the CTS, and enforce the encoded policy.
\end{itemize}
The primary limitation of this traditional architecture is that the TAS relies on a static, predefined database mapping known tasks to their required permissions. This model breaks down in the face of emergent agent behavior, which generates novel tasks not present in any database.

\section{Adapting TBAC for Emergent Tasks with an LLM Judge}
To overcome the limitations of static TBAC, we extend the model by replacing the predefined task-policy database with a dynamic, intelligent component: a Large Language Model (LLM) Judge, situated within the TAS.

\subsection{The LLM as a Policy Synthesizer}
When an agent needs to perform a novel task, instead of looking up a policy, the TAS invokes the LLM Judge to synthesize one in real time. The workflow is as follows:
\begin{enumerate}
    \item \textbf{Task Intent Submission:} An agent submits its high-level goal (e.g., "Debug the performance issue in the user-authentication service") to the TAS.
    \item \textbf{Policy Synthesis:} The LLM Judge receives the goal, along with a manifest of available tools and a set of immutable security principles (e.g., "Always prefer read-only access"). It reasons about the required steps and generates a just-in-time policy $\Pi$.
    \item \textbf{Token Issuance:} The TAS validates the syntactic correctness of the policy and instructs the CTS to issue a capability token encoding $\Pi$.
\end{enumerate}

\subsection{New Risks and Challenges}
This architecture enables dynamic authorization but introduces new, significant risks:
\begin{itemize}
    \item \textbf{Lack of Risk Awareness:} The LLM may generate a functionally correct plan that involves unnecessarily risky operations (e.g., requesting write access to a production database for a read-only task) without understanding the security implications.
    \item \textbf{Model Fallibility:} LLMs can "hallucinate" or generate flawed logic, especially for complex or out-of-distribution requests. A "naive" LLM judge has no mechanism to recognize its own knowledge gaps, potentially leading it to confidently approve a dangerous policy.
    \item \textbf{Prompt Injection Vulnerability:} A malicious agent could craft its task description to manipulate the LLM into generating an over-privileged policy.
\end{itemize}
Addressing the first two challenges—risk awareness and model fallibility—is the primary focus of our proposed extension.

\section{Uncertainty-Aware and Risk-Adaptive Policy Synthesis}
To address these new challenges, we introduce our primary contribution: an enhanced framework where the LLM Judge is made both risk-aware and self-aware of its own uncertainty.

\subsection{Formal Model Extension}
We augment the foundational model with formalisms for risk and uncertainty.
\begin{itemize}
    \item We define a static risk function, $\rho: S_{tool} \rightarrow \mathbb{R}^{+}$, which maps each tool or resource to a non-negative risk score based on its business criticality.
\end{itemize}
The LLM-generated authorization for an agent $a$ and task $t$ becomes a more comprehensive tuple, $\mathcal{A}_{a,t} = (\Pi, R_{comp}, \upsilon)$, where:
\begin{itemize}
    \item $\Pi$ is the set of permitted tool-transaction pairs (the policy).
    \item $R_{comp} = f(\{\rho(s) | s \in \tau_{tools}(t)\})$ is the composite risk score of the task, calculated by the LLM as an aggregate function (e.g., maximum or weighted sum) of the risks of all resources in the policy.
    \item $\upsilon \in [0, 1]$ is the model uncertainty estimate, representing the LLM's confidence in the generated policy $\Pi$. A value close to 1 indicates high uncertainty.
\end{itemize}

\subsection{Deep Dive: Risk and Uncertainty Components}
\subsubsection{Composite Risk ($R_{comp}$)} The static risk score $\rho(s)$ for each tool can be sourced from enterprise systems like a Configuration Management Database (CMDB), data classification labels (e.g., Public, Internal, Confidential), or manually assigned based on business impact analysis. The composite function $f$ is critical; using the maximum risk of any single tool ($f = \max(\{\rho(s)\})$) is a simple, conservative choice, ensuring that a task touching even one high-risk system is treated as high-risk overall.

\subsubsection{Model Uncertainty ($\upsilon$)} Estimating epistemic uncertainty for an LLM is an active area of research. Practical methods include:
\begin{itemize}
    \item \textbf{MC Dropout:} If the LLM architecture allows, running inference multiple times with dropout enabled and measuring the variance in the outputs, as proposed by Gal and Ghahramani \cite{gal2016dropout}.
    \item \textbf{Ensemble Methods:} Querying a small ensemble of diverse LLM judges and measuring the disagreement in their proposed policies.
    \item \textbf{Stochastic Outputs:} For a single model, performing multiple inference passes with a non-zero temperature setting and calculating the semantic variance of the generated plans.
\end{itemize}
The chosen method provides a quantifiable measure of the model's confidence in its own plan.

\subsection{Architecture of the Enhanced LLM Judge}
The enhanced LLM Judge executes an enriched, multi-stage workflow:
\begin{enumerate}
    \item \textbf{Contextual Prompting:} The TAS constructs a detailed prompt containing the agent's goal, security principles, and a \textbf{Risk-Enriched Tool Manifest}. This manifest now includes the static risk score $\rho(s)$ for each tool and its available transactions.
    \item \textbf{LLM Reasoning and Multi-faceted Output:} The LLM is instructed to perform three tasks in a chain-of-thought manner:
        \begin{itemize}
            \item First, synthesize the policy $\Pi$ to achieve the goal.
            \item Second, based on the tools in $\Pi$ and their scores from the manifest, calculate the composite risk score $R_{comp}$.
            \item Third, reflect on the complexity and novelty of the request to estimate its own model uncertainty $\upsilon$.
        \end{itemize}
    \item \textbf{Dynamic Decision and Escalation:} A central decision engine evaluates the LLM's output against configurable thresholds. The logic is:
    
        \textbf{IF} ($R_{comp} > \theta_{risk}$) \textbf{OR} ($\upsilon > \theta_{uncertainty}$) \textbf{THEN}
        \begin{itemize}
            \item Escalate to a human security officer for review and approval.
        \end{itemize}
        \textbf{ELSE}
        \begin{itemize}
            \item Autonomously approve the request and instruct the CTS to mint a token.
        \end{itemize}
    \item \textbf{Immutable Auditing:} The entire tuple—the goal, the generated policy $\Pi$, the risk score $R_{comp}$, and the uncertainty $\upsilon$—is cryptographically logged for full auditability and future model training.
\end{enumerate}

\begin{figure}
    \centering
    \includegraphics[width=0.75\linewidth]{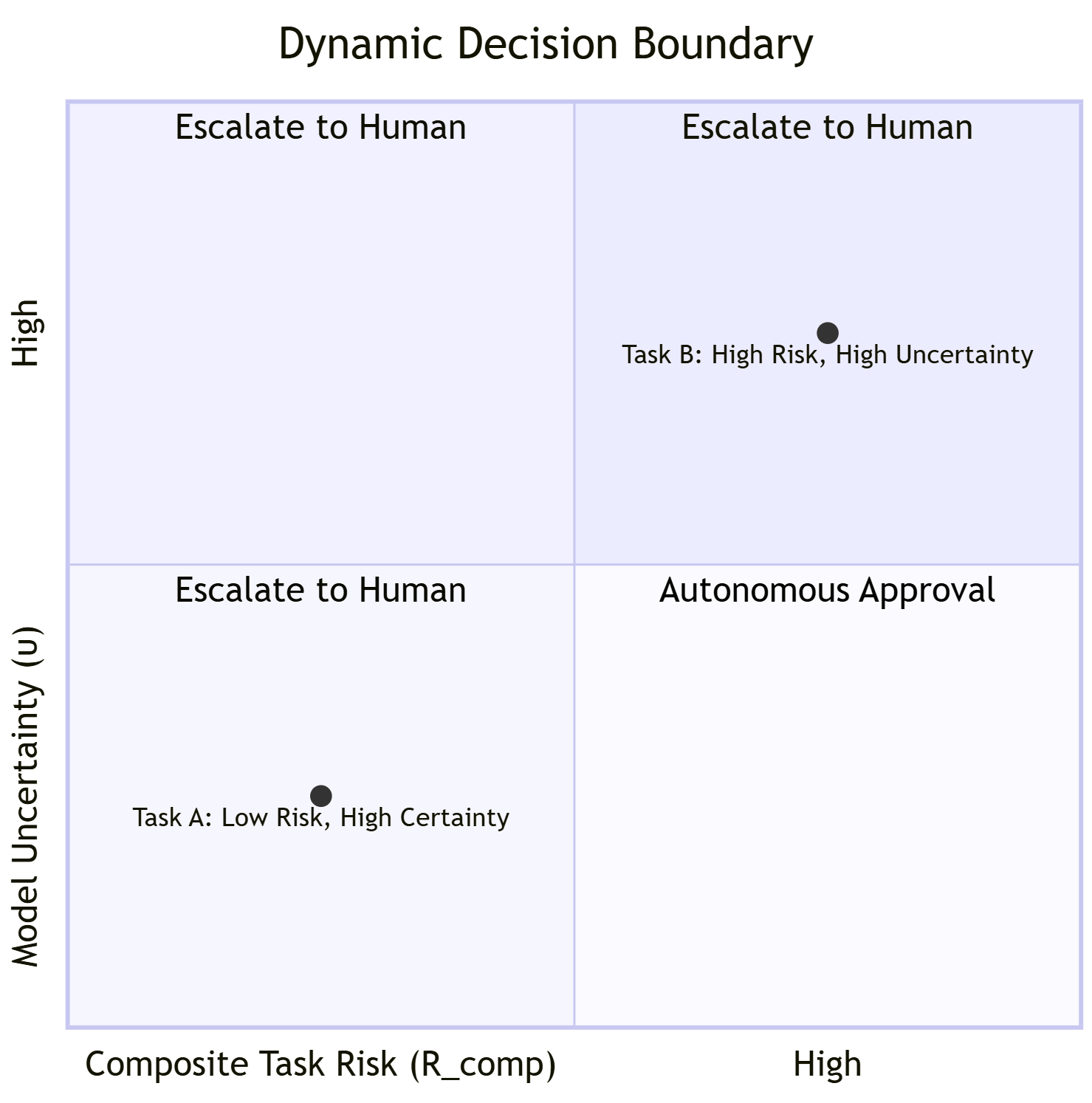}
    \caption{Decision boundaries for the LLM Judge. Requests are auto-approved only if both composite risk and model uncertainty are below their respective thresholds.}
    \label{fig:decision_boundaries}
\end{figure}

\section{Use-Case Analysis}
We analyze two contrasting scenarios to illustrate the model's behavior.

\subsection{Use Case 1: High-Risk, High-Uncertainty Incident Response}
Consider an incident response agent, `sec-agent-01`, detecting a novel threat.

\textbf{Execution Flow:}
\begin{enumerate}
\item The agent formulates the task, $T_{response}$: "Goal: Isolate compromised database `db-prod-123`, analyze its outbound traffic by mirroring traffic to a sandbox, and provision a replacement from the latest backup."
\item It submits this to the TAS. The enhanced LLM Judge is invoked.
\item \textbf{Enhanced LLM Reasoning:} The LLM consults the risk-enriched manifest. It sees that `networkAPI.updateFirewallRule()` has a risk score $\rho=9.5/10$ and `dbAPI.restoreFromBackup()` has a score $\rho=7.0/10$.
    \begin{itemize}
        \item It generates the policy $\Pi$ containing permissions for both APIs.
        \item It calculates a high composite risk score, $R_{comp} = \max(9.5, 7.0) = 9.5$, due to the critical network change.
        \item As this is a novel, complex, and high-impact scenario, its internal confidence is low, resulting in a high uncertainty estimate, $\upsilon = 0.75$.
    \end{itemize}
\item \textbf{Dynamic Decision:} The decision engine checks against thresholds, say $\theta_{risk} = 8.0$ and $\theta_{uncertainty} = 0.6$. Since both $R_{comp} > \theta_{risk}$ and $\upsilon > \theta_{uncertainty}$, the request is automatically flagged and escalated to a human security analyst's dashboard.
\item The analyst reviews the LLM's complete plan, its risk assessment, and its stated uncertainty. Agreeing with the plan, the analyst provides a one-time approval.
\item The TAS now instructs the CTS to issue a very short-lived token (e.g., 10 minutes) with heavy auditing enabled. The agent proceeds under strict supervision.
\end{enumerate}

\subsection{Use Case 2: Low-Risk, Low-Uncertainty Business Analytics}
Consider a business intelligence agent, `bi-agent-04`, performing a routine task.

\textbf{Execution Flow:}
\begin{enumerate}
\item The agent formulates the task, $T_{report}$: "Goal: Read the daily new leads from the `sales-crm` system and post a summary count to the `\#sales-updates` Slack channel."
\item It submits this to the TAS. The LLM Judge is invoked.
\item \textbf{Enhanced LLM Reasoning:} The LLM consults the manifest. It sees `crmAPI.readLeads()` has $\rho=3.0/10$ and `slackAPI.postMessage()` has $\rho=2.0/10$.
    \begin{itemize}
        \item It generates the policy $\Pi$ with read-only CRM access and Slack posting rights.
        \item It calculates a low composite risk score, $R_{comp} = \max(3.0, 2.0) = 3.0$.
        \item This is a very common and straightforward request, so the model is highly confident, resulting in a low uncertainty estimate, $\upsilon = 0.10$.
    \end{itemize}
\item \textbf{Dynamic Decision:} The decision engine checks against the same thresholds ($\theta_{risk} = 8.0, \theta_{uncertainty} = 0.6$). Since $R_{comp} < \theta_{risk}$ and $\upsilon < \theta_{uncertainty}$, the request falls into the "Auto-Approve" quadrant.
\item The TAS autonomously approves the request and instructs the CTS to issue a standard-duration token. The agent completes its task without human intervention.
\end{enumerate}

\section{Implementation Considerations and Discussion}
Deploying this model in a production environment requires addressing several practical challenges.
\begin{itemize}
    \item \textbf{Scalability and Latency:} LLM inference is computationally expensive and can introduce latency into the authorization path. For time-sensitive operations, this could be prohibitive. A potential mitigation is to implement a caching layer that stores the full authorization tuple ($\Pi, R_{comp}, \upsilon$) for previously seen, auto-approved tasks, bypassing the LLM for identical future requests.
    \item \textbf{Tool Manifest Management:} The accuracy of the risk scores in the tool manifest is paramount. This requires tight integration with enterprise asset management and data governance tools (like a CMDB) to ensure risk scores are automatically updated as systems are provisioned, patched, or have their data classification changed.
    \item \textbf{Human-in-the-Loop Workflow:} The escalation dashboard for security analysts must be designed to prevent alert fatigue. It should clearly present the agent's goal, the LLM's proposed plan, and critically, the *reasons* for escalation (e.g., "High Risk: Touches `firewallAPI`") and the model's stated uncertainty. This allows for rapid, informed decisions.
\end{itemize}

\section{Conclusion and Future Work}

The Uncertainty-Aware, Risk-Adaptive TBAC model provides a robust and trustworthy framework for securing autonomous AI agents. By making the LLM Judge explicitly reason about resource risk and its own uncertainty, we transform the access control system from a simple gatekeeper into a sophisticated risk management engine. This allows for a safer balance between agent autonomy and enterprise security.

Key areas for future work are critical for maturing this model:
\begin{itemize}
\item \textbf{Calibrating Uncertainty:} Further research into methods to ensure the LLM's uncertainty estimates are well-calibrated, meaning its stated confidence accurately reflects the true probability of its plan being correct and safe.
\item \textbf{Dynamic Risk Assessment:} Evolving the static risk function $\rho$ into a dynamic one that can update resource risk scores in real-time based on threat intelligence feeds, active security incidents, or anomalous system behavior.
\item \textbf{LLM Security and Robustness:} Developing strong defenses against prompt injection attacks specifically aimed at manipulating the LLM's risk or uncertainty calculations. This could involve using separate, specialized LLMs for plan generation versus risk assessment.
\item \textbf{Explainable AI (XAI) for Auditing:} Enhancing the audit trail with clear, human-understandable explanations for *why* the LLM arrived at a particular risk score or uncertainty level, potentially by forcing the model to output its chain-of-thought reasoning for every decision.
\end{itemize}

As agentic systems become more integrated into critical workflows, a security paradigm that is self-aware of its own limitations will be paramount. This model offers a foundational step in that direction.

\bibliographystyle{IEEEtran}
\bibliography{\jobname}

\end{document}